# Diálogos entre a Astrobiologia e a Pedagogia da Terra: Uma proposta e contribuição para a formação continuada de professores de ciências

Dialogues between Astrobiology and Earth Pedagogy: a proposal to the continued training of science teachers

Vitória Cássia Gabriela de Oliveira[1], José Alberto Casto Nogales Vera[2]
**E-mail:** vitoriaoliveirabio@gmail.com, jnogales@ufla.br.

**Abstract: Introduction and Objective.** Assuming the need to rethink Education and its function as an element of social transformation and generator of a profound ethical change and promoter of other ways of being on planet Earth, this study seeks to understand the possible dialogues between Earth Pedagogy and Astrobiology in the construction and proposition of continuing education processes for science teachers. **Methodology.** A mini-course was promoted aimed at teachers and an attempt was made to report the construction process of the training process and, using Content Analysis, to analyze the contributions pointed out by the participants to their training and teaching practice. **Results and Conclusion.** The dialogue between knowledge in the continuing education of teachers contributes with fundamental elements to teaching practice, covering cognitive, emotional, individual and collective aspects, as well as highlighting the role of the university in promoting initiatives that bring Education and Complexity together.

**Resumo: Introdução e Objetivo.** Assumindo a necessidade de repensar a Educação e a sua função como elemento de transformação social e gerador de uma profunda mudança ética e promotora de outras formas de ser/estar no planeta Terra, o presente estudo busca compreender os possíveis diálogos entre a Pedagogia da Terra e a Astrobiologia na construção e proposição de processos de formação continuada de professores de ciências. **Metodologia.** Com essa finalidade, promoveu-se um minicurso direcionado a professores. Descreve-se aqui o processo de construção do processo formativo e, valendo-se da Análise de conteúdo, analisam-se as contribuições apontadas pelos participantes a sua formação e prática docente. **Resultados e Conclusão.** O diálogo entre os saberes na formação continuada de professores contribui com elementos fundamentais à prática docente, contemplando aspectos cognitivos, emocionais, individuais e coletivos, bem como destacando o papel da universidade na promoção de iniciativas que aproximem Educação e Complexidade.
**Palavras-chave:** Formação de professores, Ensino de ciências, Educação, Complexidade.

## Introdução

O contexto no qual esta pesquisa se delineou foi repleto de desafios e mudanças profundas na maneira de educar e no próprio funcionamento da escola. Em decorrência da pandemia de Covid-19, responsável pela morte de mais de 700.000 brasileiros (Ministério da Saúde, 2023), os processos educativos em nosso país passaram por uma readaptação para o modelo remoto, que acabou por criar muitos desafios e evidenciar aspectos que precisam ser repensados no modelo educativo. O trabalho desenvolvido pelos professores tornou-se ainda mais complexo, não apenas durante a pandemia, mas também no retorno ao ensino presencial, destacando o quanto sua prática perpassa diversos saberes a serem construídos durante sua formação e trajetória profissional. Para exercer sua função social, há a necessidade de conhecer e dominar o conteúdo a ser ensinado, conhecer estratégias e teorias pedagógicas e, além disso, reconhecer aspectos históricos, sociais e culturais que constroem os sujeitos para os quais se direcionam suas aulas. É preciso considerar as experiências cotidianas na prática educativa e os elementos subjetivos que se desdobram em sua atuação política (Diniz-Pereira, 2014). Nota-se que a atividade docente não é neutra, ainda que desencorajada, sobretudo no que tange ao alicerce de uma sociedade futura, justa e igualitária.

---

[1] Universidade Federal de Lavras, Programa de Pós-Graduação em Educação Científica e Ambiental, Lavras, MG, Brasil.

[2] Universidade Federal de Lavras, Instituto de Ciências Naturais, Departamento de Física, Lavras, MG, Brasil.

A prática docente é, portanto, complexa e multifacetada, permeada por elementos que tangem desde sua formação escolar, inicial e continuada, cultural e humana (Tardif, 2002). Mais do que isso, é notório o movimento de precarização de seu trabalho, com baixos salários, grandes cargas horárias a serem cumpridas, ausência de material didático e um imergir destes profissionais em burocracias (Gadotti, 2013). Além disso, a maneira como a própria Universidade se organiza e forma professores ainda segue uma lógica "bancária", conteudista e fragmentada, sendo priorizados o ensino do conteúdo a ser lecionado em detrimento de conhecimentos pedagógicos. Diniz- Pereira (2014) classifica este modelo como promotor de uma racionalidade técnica, que tem alicerçado grande parte da formação inicial de professores em nosso país. É indiscutível que a apropriação dos saberes já postulados em sua área de formação é de fundamental importância para a atuação docente, entretanto, neste trabalho consideramos a necessidade de rompermos perspectivas conservadoras, conteudistas que fortalecem o modelo societário atual e, intencionar, lutar, propor uma formação crítica a esses professores. Justamente nesse sentido, os processos de formação continuada precisam ser pensados e construídos tanto para suprir as demandas apresentadas pelos professores em sua prática cotidiana, quanto para uma formação baseada em racionalidade crítica (Diniz-Pereira, 2014).

Muitos estudos têm evidenciado o sucesso de processos de formação continuada baseados em temas, sobretudo aqueles que possam levar para a escola outros saberes e outras maneiras de aprender e ensinar. Diante da importância e dos desafios enfrentados nesses processos formativos, construiu-se o minicurso abordando, além de perspectivas conteudistas com relação às áreas de interesse, discussões pedagógicas que perpassam e incluíssem seus papéis como educadores ambientais, contemplando também elementos e caminhos para que estes saberes fossem levados à sala de aula pelos professores. Além disso, Oliveira & Francelino (2021) Chefer & Oliveira (2022) e Gadotti (2013) muito discutem sobre a lacuna que existe na formação de professores nas áreas às quais este estudo se debruça, sendo apontado então como uma interessante lacuna de pesquisa. A proposta do curso, que além de trazer um elemento fundamental ao ensino de ciências, a curiosidade, também abraça uma teoria pedagógica que se ancora na Complexidade e em elementos fundamentais para a construção de uma nova realidade sociocultural no planeta. Muitos estudos têm evidenciado as potencialidades da Astrobiologia sobretudo no sentido de contribuir com práticas de alfabetização científica desenvolvendo a criatividade, a criticidade e, além disso, promover práticas e reflexões que possam ampliar os olhares dos discentes/docentes da relação homem/natureza/cosmo e na formação de uma identidade planetária (Oliveira & Francelino, 2021, Chefer & Oliveira, 2022). Em diálogo, a Ecopedagogia como modelo pedagógico abrange aquilo que nos parece mais caro quando pensamos uma Educação para o futuro, que construa um verdadeiro espírito e sentido de humanidade, desde a sustentabilidade até a percepção crítica da realidade.

Logo, é fundamental investir e proporcionar aos docentes acesso a cursos de formação continuada, incentivos para o contato com a prática científica em si, para que estes possam utilizar recursos que vão muito além do livro didático em suas aulas (Chefer & Oliveira, 2022). Em síntese, a proposta de pesquisa aqui construída intenciona aproximar uma ciência que se vale do devir, do movimento, a Astrobiologia e suas potencialidades, como um caminho para atender demandas apresentadas na formação continuada de professores de ciências sobretudo dentro de perspectivas que possibilitem uma educação para a Pedagogia da Terra. Neste estudo objetiva-se, portanto, relatar os elementos envolvidos na construção da proposta curricular para o minicurso, aspectos da organização e produção do evento, bem como analisar os resultados e contribuições deste na formação continuada de professores de Ciências.

**Percurso metodológico**

Este estudo caracteriza-se como qualitativo, um estudo de caso, e metodologicamente dividiu-se em duas etapas, de abordagem descritiva e exploratória respectivamente. A primeira etapa consistiu em descrever elementos que fundamentaram a construção do minicurso, desde aspectos teóricos-metodológicos até os logísticos e de organização. Esta etapa foi fundamental para delinear o contexto e bases teóricas as quais este estudo se filia. O ambiente de investigação é um minicurso pensado e construído para promover uma formação continuada de professores de ciências, dentro de perspectivas da Educação Ambiental, promovido pelo projeto "Magia da Física" da Universidade Federal de Lavras (UFLA), intitulado: *Astrobiologia e Ecopedagogia: Contribuições para o ensino de ciências.*

O evento aconteceu entre os dias 16 e 23 de novembro de 2021, de maneira remota via *Google Meet*. No formulário de pré-inscrição, 120 pessoas manifestaram interesse em participar do curso, incluindo alunos de graduação em licenciaturas. Diante desta demanda, ainda que o público-alvo fossem professores que já concluíram a formação inicial, abriu-se espaço para a participação dos mesmos, até completarmos 80 vagas de inscrições formais na plataforma SIG-UFLA. Ou seja, dos 120 pré-inscritos selecionou-se 80, priorizando professores atuantes na educação básica, mas abrindo espaço também para licenciandos. Todos os encontros foram gravados e os participantes foram convidados a responderem uma sequência de 5 questionários, distribuídos de forma online via *Google Forms*[3].

A segunda etapa metodológica deste estudo, consistiu na análise das respostas dos professores em formação continuada, a um dos questionários aplicados, o Questionário final, no qual foram recebidas 33 devolutivas do público mencionado, e que se preocupava em captar informações com relação às contribuições do minicurso em sua formação. Reforça-se que apesar da participação de alunos da graduação, utilizar-se-á como corpus de pesquisa, apenas os dados fornecidos por professores já atuantes na educação básica, nos mais diversos níveis e disciplinas. Para a análise dos dados coletados, recorremos à Análise de conteúdo, proposta por Bardin (2016) entretanto, seguindo as abordagens

---

[3] Ressalta-se que para este trabalho analisaremos apenas um dos cinco questionários.

direcionadas a pesquisas qualitativas propostas por Moraes (1999) e Minayo *et al.* (2016).

Durante as etapas de pré-análise e exploração do material, selecionamos as respostas e comentários de professores que já atuavam na educação básica e que construíram comentários complexos que ultrapassassem respostas objetivas como "sim" e "não". Sendo assim, as respostas de 9 professores (das trinta e três obtidas ao todo no questionário em questão) atenderam a estes critérios. Ainda durante a etapa de preparação do material, encontramos nos comentários dos professores aspectos que contemplavam contribuições em sua formação acadêmica/profissional, pessoal, elementos que evocavam o papel da Universidade na continuidade de processos formativos e que também se relacionavam com a realidade e cotidiano do professor em sala de aula.

Diante disso definiu-se a utilização de unidades de registros temáticas (Moraes, 1999). Em seguida, utilizando aspectos semânticos contidos nas falas, e partindo dos elementos que contemplam os saberes docentes fundamentais à sua formação profissional, propostos por Tardif (2005)[4], criamos quatro categorias (*à posteriori*), o que segundo Moraes (1999) caracteriza-se como uma análise indutiva. Para a discussão dos dados, o que corresponde às etapas finais da Análise de Conteúdo (descrição, interpretação e inferências), utilizamos como base teórica autores relevantes para esse estudo, além de Tardif (2005), como Diniz-Pereira (2014), Gadotti (2000), Chefer & Oliveira (2022) e Oliveira & Francelino (2021). Consideramos também, aspectos que se incluem nos saberes dos professores, mas que se relacionam diretamente com processos de formação continuada e o cotidiano do professor. Por fim, para evitar sua descaracterização como sujeitos de pesquisa e manter sua privacidade, conforme preconizam os procedimentos éticos de pesquisas com seres humanos, criamos nomes fictícios para cada um deles.

**Resultados e discussão**

**A construção do evento**

Inicialmente cabe destacar o caminho teórico percorrido por estes autores para a construção dessa proposta de minicurso de formação de professores. Vários foram os autores que corroboraram, direta ou indiretamente, com a sequência de conteúdos abordados, bem como com as perspectivas que extrapolavam o conteúdo científico, e alcançavam elementos fundamentais à formação crítica de sujeitos capazes de construir, ou de refletir, sobre o desenvolvimento de novas maneiras de ser/estar na Terra. destaca-se, a priori, Morin (2000), que defende a necessidade de se construir uma nova maneira de compreender o mundo, uma nova racionalidade, que supere perspectivas cartesianas, fragmentadoras e deterministas da realidade, assumindo-a como complexa. Para o sociólogo, esta mudança se reflete diretamente na maneira como. a Educação se organiza e começa a partir dos professores. Entretanto, muitos são os desafios a serem superados, visto que esses profissionais, tem suas formações iniciais e continuadas modeladas por uma racionalidade técnica (Diniz-Pereira, 2014). Para superá-la, autores como Diniz-Pereira (2014) vão defender processos formativos construídos influenciados por uma racionalidade crítica, ou seja, aquela que contemple elementos econômicos, sociais, históricos que resultaram na maneira como a sociedade se organiza, de forma a transformar e instrumentalizar os sujeitos para promover uma transformação social profunda e a formação de uma sociedade justa e equitativa. Alguns estudos vão defender como um caminho interessante para processos de formação continuada a abordagem de temas, sobretudo aqueles que possam levar para a escola outros saberes e outras maneiras de aprender e ensinar. Para além, estes temas precisam ser abordados a partir de um olhar, uma ideologia ou referencial teórico crítico. Ou seja, mais que promover a construção de saberes técnicos, estes processos precisam ser permeados da realidade sócio-histórica da humanidade e da discussão profunda de elementos que fomentam a alienação, a opressão, a injustiça e a desigualdade em nossa sociedade.

---

[4] Segundo o autor, a formação do professor, seja ela inicial ou continuada, perpassa saberes pessoais, disciplinares (com relação ao conteúdo que se ensina), pedagógicos (que incluem teorias educativas), curriculares (as normativas de ensino nacionais) e saberes da experiência (adquiridos durante a própria prática).

Nesse sentido, justifica-se a Astrobiologia que é uma ciência multidisciplinar e tem seus conceitos e lacuna de pesquisa fundamentados em teorias e trabalhos produzidos por ciências de base como a Astronomia, Cosmologia, Geologia, Astroquímica, Astrofísica dentre tantas outras (Galante *et al.*, 2016). Para compreender se realmente há a possibilidade de existir vida fora do planeta, os astrobiólogos baseiam-se em alguns princípios fundamentais: a universalidade da física e da química e como esses parâmetros se relacionam e geraram o único modelo de vida que conhecem: o terrestre. Sendo assim, para compreender a Astrobiologia, como ciência, é fundamental perpassar pela história do Universo, sua gênese, o surgimento das forças físicas, das primeiras galáxias e estrelas, dos elementos químicos, dos sistemas planetários, da Terra e, por fim, da vida.

Entretanto, além de perspectivas físico-químicas, elementos que permeiam uma abordagem conteudista e que reforçaria a racionalidade técnica, seria fundamental trazer aspectos relacionados a como todos esses processos são elucidados e descritos a partir de um olhar humano, estimulado pela curiosidade de compreender o seu papel, o sentido de sua existência e como a sua vida se relaciona com tudo o que existe no céu noturno. Considerando as preocupações socioculturais, históricas econômicas e ambientais do modelo pedagógico ao qual este estudo se filia, seria fundamental também incluir dentro da proposta elementos que contemplam a relação que se estabelece entre humanidade e Cosmos, retomando aspectos históricos, sociais que permeiam a profundidade desta ligação. A própria sequência cronológica dos eventos cosmológicos que sucedem o Big Bang já demonstra uma relação interessante de causa e efeito entre estes e como desembocaram no nascimento e evolução da vida terrestre.

A partir disso, destacam-se as contribuições de Leonardo Boff (1999), autor de grande relevância para este estudo e para a divulgação e difusão dos valores que permeiam a Carta da Terra e, por consequência, a Pedagogia da Terra: como o cuidado, o respeito e a harmonia entre todos os seres vivos que coabitam o planeta. Fato é que Boff (1999) humaniza a história do Universo, dividindo-a em cinco atos, que se inicia com o Big Bang, perpassando o surgimento das primeiras estrelas e dos elementos químicos, sua dispersão a partir das supernovas e nascimentos de novas estrelas/planeta, para chegar à vida. Para o autor a lógica sequencial da história cósmica que se mistura, ou resulta, na história da vida terrestre e caso exista, "extraterrestre", pode produzir um processo de reconexão ao Cosmos como já é comum nas culturas ancestrais.

Logo, para a proposição dos temas que foram discutidos e abordados durante o evento, a revisão de literatura foi fundamental sobretudo por possibilitar a compreensão do caminho teórico percorrido pela Astrobiologia como ciência e as necessidades contempladas em processos de formação continuada de professores que incluíssem valores, perspectivas da Pedagogia da Terra como elemento determinante na relação homem-planeta. Isso porque a Pedagogia da Terra contempla a transformação social e as maneiras de se ser/estar na Terra, constituindo-se de mais que um modelo educativo, na proposta da construção de uma sociedade alternativa global, com princípios calcados no respeito e o cuidado pelo planeta e por todos os seres vivos que coabitam, como uma única comunidade (Gadotti, 2000). Fundamental destacar também a importância do contato prévio com a Astrobiologia e com a Pedagogia da Terra em trabalhos de divulgação científica e sensibilização ambiental a partir dos conhecimentos desenvolvidos na cidade de Barbacena (Minas Gerais) desde o ano de 2019. São também frutos desse trabalho as considerações trazidas em Oliveira & Francelino (2021) acerca dos possíveis caminhos para que a Astrobiologia fosse abordada dentro de um viés ecopedagógico no ensino de ciências. Elas muito contribuem sobretudo por associar dois elementos estruturantes do trabalho.

A sequência de temas para cada um dos encontros foi construída, amparada pelos elementos discutidos até aqui, mas também considerando critérios que são discutidos neste tópico. O primeiro deles (1) consistiu em estabelecer uma sequência cronológica da história do Universo e da vida a partir do que a Ciência propõe; (2) temas que tivessem relevância dentro dos parâmetros nacionais de educação; (3) que fossem capazes de gerar discussões e percepções ambientais compatíveis com os fundamentos da Pedagogia da Terra. A Tabela 1 contém de forma sequencial os títulos e datas de cada um dos encontros, bem como os prelecionistas responsáveis pelas discussões. Ressalta-se que, *a posteriori*, partir-se-á para a caracterização de cada um dos encontros.

**Tabela 1:** Sequência de encontros do minicurso. **Fonte:** Os autores, 2023

| DATA | TÍTULO DO ENCONTRO | PRELECIONISTA |
| --- | --- | --- |
| 16/11/2021 | "Somos poeira de estrelas?" | Alexey Dodsworth |
| 17/11/2021 | "A Terra é Rara?" | Augusto Nobre Gonçalves |
| 18/11/2021 | "De onde viemos? Quem somos?" | Amanda Gonçalves Bendia |
| 19/11/2021 | "Estamos sós no Universo?" | Gabriel Gonçalves Silva e Raquel Farias |
| 22/11/2021 | Um novo olhar para a Terra e para o Universo: A Ecopedagogia | Delton Mendes Francelino |
| 23/11/2021 | Avaliação do minicurso e discussões finais. | - |

Nota-se que optamos por trazer para o evento um caráter lúdico e catártico, que pudesse instigar a curiosidade dos professores, como um elemento fundamental na produção de saberes e do ensino de ciências. Os prelecionistas responsáveis por conduzir cada um dos temas possuíam uma formação e atuação como pesquisadores em diversos campos científicos, valorizando a interdisciplinaridade tão cara a Pedagogia da Terra, e foram previamente selecionados e contatados via e-mail e, posteriormente, *Google Meet*. Todos terão seus currículos e atuação profissional descritos ao longo do texto, entretanto, cabe mencionar que todos são pesquisadores atuantes nas áreas em que foram convidados a apresentar, o que abre espaço em nosso trabalho para discutir o profundo diálogo construído entre ambientes acadêmicos e a escola básica, pesquisadores e professores, e o papel fundamental da Universidade pública no estabelecimento deste contato e na formação dos professores. Mais que isso, defende-se aqui que esta articulação é um elemento fundamental à construção de uma educação crítica e constituidor da escola cidadã.

Iniciamos o minicurso no dia 16 de novembro de 2021, às 19 horas com o encontro, intitulado "*Somos poeira de estrelas?"*, que se dedicou a trazer discussões acerca da origem do Universo, a partir da teoria mais aceita na atualidade, a Teoria do Big Bang, perpassando elementos que constituem a maneira como a humanidade entendeu e interpretou o cosmos, a partir da habilidade de observação de padrões, fundamentais ao senso comum e posteriormente, as visões mitológicas, o nascimento da filosofia e como a Ciência resulta de todo esse percurso epistemológico (Galante *et al.*, 2016). Para isso convidamos o Prof. Dr. Alexey Dodsworth, que é Doutor em Filosofia, além de cursar nova pós-graduação na área de Ensino de Astronomia. Nota-se que são elencadas aqui perspectivas fundamentais à compreensão da natureza da Ciência, de como ela se firma como uma prática social e humana, permeada por muitas controvérsias, aspectos que, segundo Pérez *et al.* (2001) interferem diretamente na maneira como os professores a ensinam.

O título do encontro é uma famosa frase do astrônomo norte americano Carl Sagan que, em seu trabalho de divulgação de ciência, desde 1980, já trazia elementos capazes de produzir catarses no público que contempla a grandeza do Universo diante da Terra e dos humanos e, ao mesmo tempo, conceber a belíssima possibilidade de conhecê-lo e compreendê-lo segundo a evolução complexa que originou a racionalidade. O título também abriu caminhos para a aproximação dos professores com o fato de que toda a matéria, ou energia estabilizada após o Big Bang, liberada há bilhões de anos por estrelas primitivas, constitui tudo o que conhecemos, atribuímos sentidos e significados, inclusive, nossos corpos. Estes podem ser "starts" interessantes para o despertar de uma conexão com toda a história cósmica e com tudo o que existe, visão e valores que tentam ser recuperados pela Pedagogia da Terra (como por meio da consciência e cidadania planetária/cósmica) e que apesar de não anularem perspectivas científicas, podem representar um caminho interessante para ressignificar a existência humana na Terra.

No segundo encontro, que aconteceu no dia 17 de novembro de 2021, intitulado "*A Terra é rara?"*, aprofundamo-nos nos processos de formação e características terrestres que foram/são fundamentais à vida e que se ligam diretamente ao nascimento e características da estrela matriz, o Sol. A partir disso, analisamos as possibilidades de encontro de tais condições em exoplanetas. Para este dia, convidamos o Prof. Dr. Augusto Nobre Gonçalves, que é geólogo e doutor em Engenharia de Materiais e Nanotecnologia. O título do encontro foi inspirado em uma teoria proposta por dois cientistas no início do século XX e que, contrariando perspectivas de astrônomos renomados da época, propunha que a Terra fosse um planeta especial dentre os outros milhares existentes. Isso porque, apenas em nosso lar existiram/existem condições e coincidências astronômicas e geológicas que possibilitaram o surgimento da vida complexa e inteligente, algo que não se repetiria facilmente no cosmos (Blázquez & González, 2020).

Fica evidente que, com o questionamento proposto para a discussão duas visões seriam contempladas, desde a de que a Terra é um planeta que apresenta singularidades extraordinárias, estando dentro de um limiar cósmico e astronômico que possibilitou o desenvolvimento de vida e, ao mesmo tempo, quando lançamos olhos para fora daqui, encontra-se um incontável número de mundos possíveis (Galante *et al*., 2016). Gadotti (2000) nos lembra que para amar o planeta Terra, precisamos conhecê-la e compreender como as dinâmicas planetárias (fluxos de energia, de elementos químicos entre tantos outros) que possibilitam/mantém a vida são resultados de bilhões de anos de evolução planetária que independem da existência humana. Olhar para outros planetas que, ainda que similares, não seriam capazes de abrigar vida como a terrestre, também nos convoca a reconhecer a "generosidade" do planeta Terra e a repensar o papel que aqui cumprimos (Gadotti, 2000).

Já o terceiro encontro ocorreu no dia 19 de novembro de 2021 e se iniciou às 19 horas. Este dia, nomeamos *"De onde viemos? Quem somos?"* perguntas estas que são feitas pela humanidade desde seus primeiros contatos com o Cosmos. Perpassamos questões relacionadas à origem da vida no planeta Terra, elementos que constituem a visão do papel que ocupamos como espécie racional neste contexto Universal e terrestre e debruçamo-nos na fragilidade e força da vida planetária. Nesse dia, recebemos a Prof. Doutora Amanda Gonçalves Bendia, que é Bióloga, doutora em Microbiologia e se dedica ao estudo de microrganismos extremófilos e como esses seres podem contribuir para as buscas de vida fora da Terra.

Houve a intenção de discutir o conceito de vida, ainda que existam muitas controvérsias para a definição, visto que esta implicaria em questões dos mais diversos campos, desde as ciências da natureza, até mesmo esbarrando em perspectivas das ciências humanas como a Filosofia, a sociologia, a política e a ética. Considerou-se fundamental que os processos de origem da vida aqui discutidos, baseados em Ciência, demonstrassem o efeito cascata de eventos cósmicos, no Sistema Solar e na própria Terra que possibilitaram o surgimento e manutenção da vida. Além disso, buscou-se aprofundar em elementos que demonstrassem as profundas inter-relações e interdependência entre Terra e vida, demonstrando a autopoiese da vida e o incessante devir destas relações, algo bastante discutido e defendido como elemento de sensibilização ambiental pela Pedagogia da Terra, já que o planeta e a biodiversidade compartilham um destino comum (Gutiérrez & Prado, 2013; Gadotti, 2000).

O quarto encontro, intitulado *"Estamos sozinhos no Universo?"*, foi conduzido pelos pesquisadores Gabriel Gonçalves Silva e Raquel Farias que são, respectivamente, engenheiro químico e doutorando em Química pela UFSCar (Universidade Federal de São Carlos), com foco em Astrobiologia e Meteorítica; biotecnóloga pela UFRJ (Universidade Federal do Rio Janeiro) e atualmente, doutoranda em Biotecnologia pela USP (Universidade Federal de São Paulo), onde se dedica a estudar a resistência de microalgas às condições extremas de Marte para futuros processos de exploração espacial. Neste dia, focamos a discussão na possibilidade de existência de vida fora da Terra, no Sistema Solar e fora dele. Com foco nas diferenças entre a vida microbiana (extremófilos) e vida inteligente, trazendo perspectivas reais acerca das pesquisas atuais em Astrobiologia que focam, principalmente em seres vivos basais, os extremófilos e; desconstruir visões amplamente divulgadas pela mídia, motivadas pela curiosidade em relação à vida extraterrestre.

Gadotti (2000), sobretudo por suas influências da Pedagogia freiriana, muito discute sobre a importância do encantamento para processos de ensino aprendizagem, sobretudo por este despertar a curiosidade dos educandos, provocar o questionamento, a criticidade. Estes elementos também figuram como fundamentais ao ensino de ciências, sobretudo por também possibilitar o evocar de saberes prévios dos alunos e o erigir de hipóteses. Além de este tema trazer os elementos discutidos, também era objetivo aproximar os professores da prática científica.

O quinto e penúltimo encontro *"Um novo olhar para a Terra e para o Universo: A Ecopedagogia"* teve como proposta basilar apresentar a Ecopedagogia como teoria pedagógica, seus aspectos estruturantes e, a partir dela, lançar olhares a processos de ensino de ciências e para o futuro da vida na Terra e no Universo. No campo da educação ambiental, são bastante comuns discussões acerca da transversalidade, não entrando como tema, mas como uma abordagem acerca dos conteúdos discutidos. Entretanto, como o evento era direcionado a professores que atuam como educadores e como educadores ambientais, considerou-se importante que em um dos dias de curso fossem abordados aspectos basilares da Ecopedagogia como modelo pedagógico, bem como discussões acerca de como esta poderia se encaixar em processos de ensino de ciências instigantes e investigativos.

Para este dia, convidamos o prof. Ms. Delton Mendes Francelino, que é Biólogo, professor, pesquisador, produtor cultural na cidade de Barbacena (MG) e, atualmente, doutorando na Universidade Federal de Minas Gerais. Também desenvolve diversos projetos de divulgação científica voltados para a Astronomia como campo do conhecimento fundamental para a sensibilização ambiental. Dedicou-se neste dia a discutir que a Pedagogia da Terra é um convite para o retorno de uma consciência humana com relação à Terra e à própria humanidade baseada nos princípios da integralidade e complexidade das relações vida e Terra e o nascimento de um espírito verdadeiramente humano (Morin & Kern, 2003): a formação de sujeitos coletivos. Além disso, buscou mostrar as potencialidades da Astrobiologia como eixo temático para erigir sentidos sobretudo em processos educativos de ciências em ambientes formais de ensino.

Fechamos o minicurso com o encontro do dia 23 de novembro, quando propusemos uma roda de conversa com todos os professores participantes, inclusive os prelecionistas. Este dia foi essencial para que todos pudessem compartilhar as impressões, dúvidas, ideias obtidas no curso e aprofundar ainda mais a proposta a partir disto. Acredita-se que este tenha sido um dos dias que mais contribuíram com a pesquisa, sobretudo pelo espaço de diálogo dado aos docentes, prelecionistas e até mesmo a nós, pesquisadores envolvidos. Recorremos durante toda a trajetória de construção do minicurso a aspectos teóricos que fundamentam a pesquisa e que poderiam mostrar os possíveis diálogos entre a Astrobiologia e a Pedagogia da Terra em processos de ensino de ciências para a complexidade e geradores de sujeitos críticos e transformadores sociais. Ressaltamos ainda que a frequência durante os dias do minicurso por parte dos professores foi bastante variada, mas estes tiveram acesso a todas as gravações dos encontros para acompanhar o curso na íntegra.

Neste tópico, elencamos algumas das instâncias de diálogos estabelecidas entre os dois campos em estudo, entretanto, é relevante direcionarmos os esforços para entender como os professores participantes compreenderam esta aproximação. Sendo assim, no tópico que se segue, trouxemos as análises iniciais de parte do material coletado, a fim de respondermos a esta questão. Informações sobre o minicurso, bem como os vídeos dos encontros e materiais de leitura indicados podem ser encontrados no site oficial do "Festa nas estrelas", no link: https://sites.google.com/ufla.br/festa-das-estrelas/astrobiologia?authuser=0.

**Contribuições do evento para a formação dos professores**

Neste tópico é apresentada uma discussão acerca das contribuições do minicurso na formação dos professores. O material analisado aqui, foi extraído do último questionário, aplicado no encontro final e que tinha como objetivo que os professores participantes colocassem suas opiniões, críticas e avaliassem se houve contribuição em sua formação e de que forma contribuiu. Foram obtidas 33 respostas no total, somando graduandos em licenciaturas e professores em formação continuada e os apontamentos com relação ao minicurso foram bastante positivos. Importante relembrar que, os dados fornecidos pelos estudantes da graduação, ou seja, em formação inicial foram desconsiderados. As nove respostas analisadas são de professores participantes em formação continuada, que fazem parte do grupo inicial de 33 inscritos e que se caracterizam da seguinte forma: cerca de 42% possuíam formação inicial em Ciências biológicas, seguidos por pedagogos (21%) e químicos (15%). Ainda contamos com historiadores, químicos, geólogos, matemáticos e bacharéis em letras. Cerca de 42 % também possuíam especialização, 12 % de mestres e 9% doutores. A maioria dos professores atuavam na educação básica em um intervalo de 5 a 10 anos de experiência. Muitos dos professores voluntários afirmaram que já se interessavam por Astronomia e campos científicos correlatos, como a Astrobiologia, termo que despertou bastante curiosidade. Além disso, foi apontado por alguns inscritos que estes consideravam construir projetos e propostas pedagógicas a serem utilizadas em suas aulas, a partir dos temas abordados no curso. Diante do formato online, os professores residiam em diversos estados do país, com grande destaque e alcance em Minas Gerais, sobretudo nas cidades de Barbacena, Lavras e Nepomuceno. Inicialmente, os participantes foram questionados se houve algum dos encontros que lhes foi mais surpreendente, interessante, que despertou mais curiosidade (Figura 1). Ressalta-se que para a construção da Figura 1, como se trata de dados quantitativos, foram consideradas às 33 respostas ao Questionário em análise.

**Figura 1:** O encontro mais surpreendente, segundo as respostas recebidas. **Fonte:** Os autores, 2023

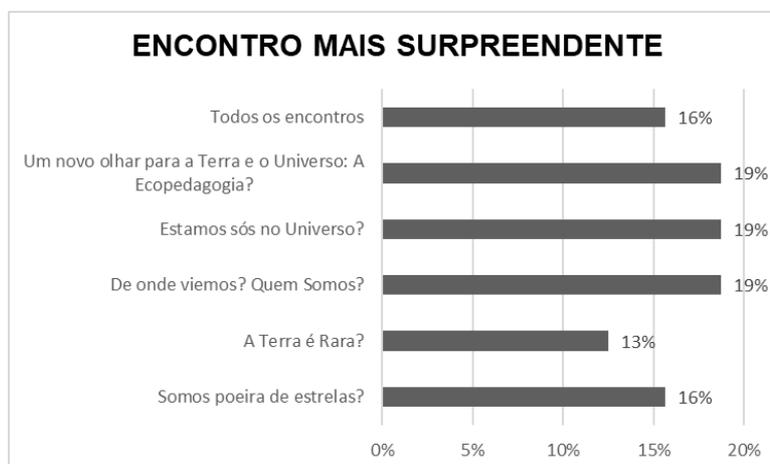

Nota-se que houve uma distribuição equilibrada entre os encontros. Entretanto, os três mais votados, quando mencionados, além de estarem diretamente ligados à assuntos que são objeto de pesquisa da Astrobiologia na atualidade, como a origem da vida, seres vivos extremófilos, e a relação Terra/vida, trouxeram à tona aspectos que se relacionavam diretamente com o cotidiano e experiências dos prelecionistas, como cientistas, o que acabou por favorecer uma proximidade com o público. Além disso, a discussão construída no encontro "Um novo olhar para a Terra e para o Universo: A Ecopedagogia", trazia um olhar humanizado e didático para a temática ambiental a partir da Astrobiologia. Esta perspectiva foi reforçada por comentários dos professores que mencionaram a "maneira como os professores explicaram o conteúdo", "a aproximação da linguagem acadêmica com os professores participantes".

Todos os professores voluntários afirmaram que o curso contribuiu para a sua formação de alguma maneira. Muitos elogiaram a qualidade dos encontros e dos materiais sugeridos para leitura. Entretanto, alguns deles teceram críticas com relação ao horário de início e fim dos encontros, já que em virtude das discussões após a palestra excedemos o período previamente estabelecido, que era das 19:00 às 21:00 horas. Nesse sentido, consideramos importante analisar de forma mais aprofundada os comentários que eles fizeram com relação às contribuições do evento. Iniciaremos com as falas das professores Ana e Tereza que estão agrupadas na primeira categoria de análise, que corresponde ao papel da Universidade na formação de professores, e descrita a seguir:

> *"As universidades deveriam ter mais cursos assim, para podermos dialogar com assuntos tão diversos do nosso cotidiano, que, porém, estão inseridos o tempo todo em nossas vidas."* **(Professora Ana)**

> *"(...) e gostei muito porque foi voltada para nós professores. isso que amei. foi um nível acadêmico para nós professores do ensino fundamental. E isso foi muito maravilhoso."* **(Professora Tereza)**

Nas falas das professoras encontramos elementos que caracterizam o papel da Universidade, não só na formação inicial quanto continuada dos professores e sua relação com a escola. Ana discorre sobre um dos aspectos que fundamentam a Universidade como instituição pública: a Extensão (função esta que abre espaço para o diálogo do que é produzido academicamente com a comunidade local, global e, em especial, as escolas). Sabe-se da importância de tais instituições na formação inicial e continuada dos professores em nosso país e, em contrapartida, na fala de Ana, entende-se que estas iniciativas (sobretudo de formação contínua) vem sendo insuficientes; ainda que com iniciativas interessantes encontradas, alguns dos artigos referenciados aqui corroboram a afirmação da mesma e vêm demonstrando e discutindo a distância entre as Universidades e as Escolas, sendo apontados como uma realidade no mundo todo (Chefer & Oliveira, 2022, Pansera de Araújo *et al.*, 2009).

Mais do que isso, entende-se que tais projetos não estão alcançando os professores, seja por falta de divulgação ou ainda os modelos de formações que inviabilizam sua participação. Nos últimos anos, diante do enfrentamento da pandemia da Covid-19, reforçou-se ainda mais a necessidade de se investir em uma formação contínua do professor, não no sentido de uma racionalidade técnica, de atualização, já que, apesar do fundamental papel das Universidades públicas, os professores também são produtores de saberes em seu cotidiano (Tardif, 2005); mas elencando elementos que os permitam o contato com saberes específicos, disciplinares, mas que também reflitam sobre o mundo, a sociedade e sua própria ação docente (Diniz-Pereira, 2014).

Como um reflexo disso, encontramos no comentário da Professora Tereza uma discussão no que se refere à distância existente e muito discutida, entre o ambiente acadêmico, sua linguagem e especificidades e a sociedade geral e, incluindo a escola e os professores. Latour (2012) é um dos autores que discutem esse distanciamento, comparando a Ciência Moderna, sua prática e produtos, à uma "caixa preta", de difícil acesso e compreensão para quem a vê de fora. Esta foi uma das principais preocupações trazidas por nós, pesquisadores e organizadores do evento, durante as reuniões de convite aos prelecionistas, à proposição de dias e horários para que não houvesse congruências com o horário de trabalho dos professores, a integração de expressões artísticas como a poesia, quadros, músicas dentro dos encontros, a construção de práticas pedagógicas pelos próprios prelecionistas com relação ao tema que abordaram e disponibilização deste material aos professores.

Caminhamos então para a segunda categoria de análise, a partir das seguintes afirmações:

> *"Depois de uma jornada de 12 horas trabalhando, mesmo cansada foi gratificante e enriquecedor o Minicurso. Uma sugestão seria que esses pudessem ser mais no início do ano, quando estamos um pouco mais tranquilos, pois essa época de fechamento de notas foi um pouco complicado para mim, até mesmo devido ao cansaço mental, mas espero ser convidada para mais eventos como esse."* **(Professora Paula)**

> *"Com o passar do tempo fui me acomodando à rotina da sala de aula. Muitas vezes repetindo a mesma coisa durante anos. Esse minicurso foi muito interessante, parece que abriu os meus horizontes. Parece que descobri outro mundo. Sinto que a Ciência é apenas uma criança, que tem muitas interrogações, muitas lacunas."* **(Professora Carla)**

As professoras Paula e Carla pontuam questões relacionadas ao seu cotidiano docente, que constituem a segunda categoria. Paula, destaca sobretudo sua sobrecarga devido a grande quantidade de horas trabalhadas, bem como, como menciona Gadotti (2013), o empilhamento de burocracias e funções que se acumulam ao professor e que o distanciam de seu papel como intelectual crítico (Teixeira *et al.*, 2017). Em contrapartida, Carla comenta que por anos, repetiu as mesmas práticas com turmas diferentes e em contextos diferentes, e que o contato com o tema gerou um incômodo, um estímulo para sair da "zona de conforto". Obviamente, esta também é uma das funções de processos de formação continuada, que seja um espaço para a revisão crítica de suas atuações, práticas pedagógicas entre outros, mas, o mais importante nesta categoria de análise é refletir sobre o que levou Carla a esse sentimento de "acomodação". Temos, em um primeiro momento, uma professora que relata sua exaustão, e em outro, uma que fala sobre ter se "acomodado". Seria esta "acomodação" fruto de uma preguiça, ou há uma causa e efeito nos comentários das professoras?

O comentário de Carla também traz em essência, algo que Pérez et al., (2001) conjecturam, a maneira a concepção de ciência do professor e os impactos disso nos discentes. Nota-se que houve uma ruptura na maneira como a professora compreendia a Ciência, de algo fechado, terminado, para algo que ainda há muito o que ser investigado, conhecido. Tal perspectiva é importante para este estudo já que, como modelo pedagógico, a Pedagogia da Terra se baseia na curiosidade, no encantamento, na criatividade e criticidade para a formação de uma outra maneira de ser/estar no mundo, de passivo para ativo (Gadotti, 2000); algo que só é possível compreendendo a natureza da Ciência, que se encontra no cotidiano, não sendo apenas obras de grandes gênios e de que ainda existem muitas lacunas possíveis. Também é neste ponto que o diálogo entre a Astrobiologia e a Pedagogia da Terra no ensino de ciências é interessante, já que a primeira é uma área em amplo crescimento, com muitas perguntas e que, por isso, consegue articular esta visão de ciência em processos educativos.

Para a terceira categoria, partir-se-á das falas dos professores Bruno, Laura e Pedro:

> *"Só tenho a agradecer pelo excelente evento e dizer ele contribuiu muito para o meu crescimento profissional visto que gosto bastante desta área de astronomia e juntando as duas coisas que eu amo que são biologia e astronomia isso é ainda melhor. Estou ansioso pelo próximo evento."* **(Professor Bruno)**

> *"Contribuiu e muito, por ser uma área que não tinha muita afinidade, me despertou um interesse muito grande em aprofundar sobre os temas. E um aprendizado que tive, o aluno pode não gostar do tema, mas a forma como lhe é apresentado pode fazer com que ele mude sua visão."* **(Professora Laura)**

> *"me surpreendeu muito o curso, não esperava ver tanta biologia em um minicurso da física, fiquei encantada."* **(Professora Luíza)**

Os professores Bruno, Laura e Luíza trouxeram elementos que se relacionam com o que Tardif (2005), chama de saberes disciplinares, ou seja, do conteúdo que ministram. Nota-se que Bruno, diferentemente dos professores Laura e Luíza já possuía um contato prévio e afinidade com Astronomia/Astrobiologia. Entretanto, um trecho da fala de Laura chama atenção *"E um aprendizado que tive, o aluno pode não gostar do tema, mas a forma como lhe é apresentado pode fazer com que ele mude sua visão"*. Segundo ela, este trecho justifica o seu encantamento com relação ao tema após o contato, trazendo elementos provenientes de sua própria experiência como discente (no curso) e docente na sala de aula. Lembro-lhes, que o encantamento é um elemento fundamental para uma Pedagogia da Terra, já que este produz um erigir de sentidos, percepções que vão além de saberes técnicos e se direcionam a formação dos indivíduos de forma integral: racional, emocional e crítica (Gadotti, 2000; Morin, 2000).

Além dos aspectos mencionados, é importante destacar que na fala dos três professores, encontramos elementos que evocam o potencial da Astrobiologia como eixo temático para uma educação menos fragmentada, conectando saberes dos mais diversos campos (desde as ciências humanas às ciências naturais), aspecto fundamental para uma compreensão da realidade de forma interligada, interconectada e complexa. Para Gadotti (2000), processos educativos direcionados a partir de visões inter/transdisciplinares são aspectos basilares de uma Educação de Futuro e para a formação de sujeitos ecológicos conscientes de seu papel/lugar (biológico, ético, social, político) no "Sistema Terra/Cosmos".

Por fim, partindo da premissa do contexto em que o educador está inserido, um sistema educativo que desconsidera as nuances do humano, valorizando a racionalidade técnica, perspectivas reducionistas e fragmentadas, construiu-se a categoria a seguir discutida, a partir das respostas das professoras Karen e Eliane:

*"Foi o melhor curso dos cursos que fiz este ano. Acho que o melhor em anos, no sentido de repensar as ideias. repensar o meio ambiente, repensar a educação."* **(Professora Karen)**

*"O curso foi de grande valia na minha formação não só como licenciada em biologia, mas também na minha formação como pessoa, me fazendo refletir conceitos novos e repensar conceitos antigos. Consegui perceber o quanto o tema astrobiologia é rico para se trabalhar em sala de aula, integralizando na prática os conhecimentos de diferentes áreas da ciência. Acredito que no geral, o curso superou minhas expectativas e me deixou um pouco mais consciente do meu papel na sociedade. Obrigada e parabéns a todos os envolvidos na organização!!"* **(Professora Eliane)**

Os comentários das professoras Karen e Eliane são riquíssimos em elementos e significados. Classificamos como "formação humana", mas perpassam algumas das discussões já tidas até aqui e importantes para esta pesquisa, já que acaba tocando no cerne da pergunta principal de pesquisa a qual este trabalho integra[5]: Os saberes da Astrobiologia, como elemento de integração de ciências e conhecimento e geração de estímulos, sentidos e consciência ambiental planetária/cósmica, como a Pedagogia da Terra propõe. Além disso, Tardif (2005) muito discute sobre a inseparabilidade do docente como sujeito individual e seu trabalho. Nos últimos anos, sobretudo com o fortalecimento de políticas neoliberais, muito tem sido discutido sobre uma educação neutra, desconectada de um viés político-ideológico, algo que, diante do exposto sobre a natureza do trabalho do professor, seria impossível. Mais do que isso, uma educação neutra é também acrítica, logo, os esforços para o silenciamento dos professores também é uma estratégia para manter perspectivas hegemônicas baseadas na exploração, na dominação e no capital.

Nesse sentido, construir propostas de formação continuada de professores que tenham influência, como mencionado pelas professoras, em suas maneiras de ser/estar no mundo e em sociedade, em repensar o movimento dos processos educativos e sociais, como se modelam e se constroem, projetando uma visão do futuro que estamos construindo são elementos fundamentais a uma racionalidade

---

[5] Destaca-se que este estudo faz parte de uma pesquisa de mestrado, conduzida e orientada pelos autores.

crítica (Diniz--Pereira, 2014) e ao que propõe a Ecopedagogia como modelo ético e educativo.

**Conclusão**

Em conclusão, os resultados evidenciaram que houve adesão e curiosidade com relação ao tema por professores das mais diversas áreas e formações iniciais. A maioria deles já havia tido contato com ciências correlatas à Astrobiologia, mas poucos com a Ecopedagogia como modelo pedagógico e ético. Além disso, o formato online e a distribuição dos horários e datas favoreceu e facilitou a participação dos professores e o contato com os cientistas/especialistas das mais diversas áreas. Como apontado em algumas falas, o contato dos participantes graduados em disciplinas diferentes, com a arte, a ciência e até mesmo imagens favoreceram um ambiente inter/transdisciplinar, ainda que sejam necessários muitos avanços neste sentido.

Percebemos a partir dos encontros aproximações entre a Astrobiologia e a Pedagogia da Terra, bem como as contribuições do minicurso e deste diálogo nos mais diversos aspectos formativos que caracterizam e são fundamentais na prática docente. As categorias de análise geradas demonstraram que a promoção de processos formativos que contemplem as nuances do "ser educador" desde elementos cognitivos e racionais, aos emocionais e intuitivos que formam uma inteligência integral: validando aspectos cotidianos, disciplinares, políticos, históricos, sociais, emocionais e racionais intrincados à sua formação humana, precisam ser pensados e construídos, sobretudo no sentido de construir uma mudança profunda de valores e atitudes antrópicas para com o planeta Terra.

Fundamental destacar o papel das Universidades Federais em nosso país na promoção deste tipo de iniciativa, cumprindo o seu papel institucional, e gerando espaços de reflexão crítica não somente das práticas de ensino aprendizagem, mas também da Educação como instrumento de humanização do humano e alicerce para a transformação social que se propõe com a Pedagogia da Terra. A luta é para que novos espaços como esses sejam abertos e que as Universidades se engajem também na promoção de uma educação para a Complexidade.

Conclui-se que a formação de professores é um processo que não se esgota na formação inicial e nem mesmo em especializações e formações continuadas; é permanente, está no cotidiano na avaliação de suas ações na sala de aula, no aprimoramento da gestão de informações e sentimentos característicos do contexto planetário atual e vai além de um posicionamento político crítico perante as mazelas que se apresentam no planeta. Ressaltamos também a necessidade de entender, em pesquisas futuras, de que forma este diálogo chega à escola e se tem potencial de geração de sentidos/afetos reflexões socioambientais, tal como ocorreu no minicurso para formação de professores.

**Agradecimentos**

**Referências**